# Analyses on Christoph Clavius' Reports of Total Solar Eclipses in 1560 and 1567: Key References for the Centennial Variations of the Earth's Rotation Speed and the Solar Radius

Hisashi Hayakawa[1] (1–4), Mitsuru Sôma (5), Jean-Pierre Rozelot (6),
Koji Murata (7–8), Alexei A. Pevtsov (9), Stanislav Gunár (10), Lucie M. Green (11)

(1) Institute for Space-Earth Environmental Research, Nagoya University, Nagoya, 4648601, Japan

(2) Institute for Advanced Research, Nagoya University, Nagoya, 4648601, Japan

(3) Space Physics and Operations Division, RAL Space, Rutherford Appleton Laboratory, Science and Technology Facilities Council, Harwell Oxford, Didcot, Oxfordshire, OX11 0QX, UK

(4) Nishina Centre, Riken, Wako, 3510198, Japan

(5) National Astronomical Observatory of Japan, Mitaka, 1818588, Japan

(6) Université de la Côte d'Azur (OCA), Grasse, 06130, France

(7) Institute of Library, Information and Media Science, University of Tsukuba, Tsukuba, 3058550, Japan

(8) Center for Digital Humanities and Social Sciences, Nagoya University, Nagoya, 4648601, Japan

(9) National Solar Observatory, Boulder, CO 80303, USA

(10) Astronomical Institute of the Czech Academy of Sciences, 25165 Ondřejov, Czech Republic

(11) Mullard Space Science Laboratory, University College London, Surrey, RH5 6NT, UK

**Abstract**

Variations in solar radius (hereafter $R_{\odot}$) is a key reference for solar magnetic activity in time. The sunlight amount may have varied with $R_{\odot}$ and had an effect on the Earth's climate in the past. Eclipse observations offer a unique opportunity to measure the absolute $R_{\odot}$ value before modern direct observations. The scientific community has discussed a possible long-term $R_{\odot}$ variability from 1715 onward. Prior to their coverage, Clavius' eclipse reports had been subjected to qualitative debates regarding the local eclipse visibility and a possible secular $R_{\odot}$ trend. This study leverages the recent dramatic developments of lunar topography data and ephemeris data to provide an effective resolution of this debate. Clavius' eclipse reports described an explicit totality in 1560 at Coimbra and a "slender circle" around the eclipsing Moon in 1567 at Rome. Our study revised the $\Delta T$ constraints of $-492$ s $\leq \Delta T \leq 200$ s in 1560 and 140 s $\leq \Delta T \leq$ 151 s in 1567 to satisfy Clavius' descriptions, considering the lunar limb profile and assuming Auwers' canonical $R_\odot$. This study constrains the $R_\odot$ margin of 1567, utilising three scenarios to interpret Clavius' account. The local totality requires an upper $R_\odot$ limit of 1567 as $R_\odot \leq$ 696200 km in absolute size (959.92" in angular size), indicating no linear secular $R_\odot$ shrinkage but possible $R_\odot$ oscillations on a centennial timescale. Conversely, the annularity scenario is considered unlikely because it requires an $R_\odot$ decrease of 7.5" within 3 centuries, even beyond the capacity of extreme shrinking-Sun hypotheses.

[1] hisashi@nagoya-u.jp

**1. Introduction**

The size of a star is an emergence property that is governed by physical processes taking place in the stellar interior. The size of our Sun has been a classic topic of interest in astronomy and astrophysics and our ability to measure the solar radius with high precision has led to it being used as a reference in other astronomical studies. Solar radius (hereafter $R_\odot$) measurements have been conducted since antiquity with numerous different methodologies, as outlined in Table 1 of the study by Rozelot and Damiani (2012). The scientific community uses Auwers' canonical $R_\odot$ value of ≈ 696000 km in absolute size (959.63" in angular size) based on measurements in the 1870s to 1880s (Auwers, 1891; Archinal *et al*., 2011). However, this value has been revised to the so-called nominal value of 695700 km (Prša *et al*., 2016) and the modern standard (helioseismic) value of 695780 km (Takata and Gough, 2024). These $R_\odot$ values have been widely used to develop a variety of studies in astronomy, astrophysics, and geophysics.

The question of how $R_\odot$ evolves over time has long been debated[2]. Eddy *et al*. (1980, hereafter E+80) raised the possibility of a steep decrease in $R_\odot$ of 0.1% per century based on meridian transit measurements from 1836 to 1953. Their results have been heavily debated in contemporaneous studies (Dunham *et al*., 1980, 2016, hereafter D+80 and D+16; Parkinson *et al*., 1980; Sofia *et al*., 1983, hereafter S+83). This question has been occasionally debated even recently, if not intensively (Pap *et al*., 2001; Sofia *et al*., 2013; Hiremath *et al*., 2020). However, the scientific community has not reached any substantial conclusions owing to technical difficulties such as the atmospheric effect on $R_{\odot}$ measurements and limited chronological coverage (RD12; Rozelot *et al*., 2018).

[2] Modern modeling of stellar atmospheres (*e.g.*, Ribas, 2009) suggests that the Sun's radius decreased from about 0.95 to about 0.9 of the current $R_\odot$, shortly after the Sun's formation (timescales of hundred million years). Since then, the $R_\odot$ was gradually increasing. While these evolutionary changes take place over a very long period of time (billion of years), there could be variations on much smaller time scales related to the level of solar activity. Thus, for example, variation in production of magnetic field may result in changes in partitioning internal energy inside the Sun that would impact solar irradiance and, perhaps, solar radius (*e.g.*, Pevtsov, 2012).

Eclipse observations offer a unique opportunity to measure and visualise the Earth's rotation speed, $R_{\odot}$, and solar coronal structure, minimising these atmospheric effects and maximising the chronological coverage (RD12; D+16; Pasachoff, 2017). Owing to their spectacular significance, total solar eclipses have been documented in human history since the late 8th century BCE, including datable reports that allowed for occasional reconstructions of these variations in the past (RD12; D+16; Stephenson *et al*., 2016; Hayakawa *et al*., 2025). Historical eclipse records have particularly contributed to measurements of the Earth's rotation speed in the past, based on the relative locations of the observational sites and the totality/annularity path for each case of total and annular solar eclipses (hereafter TSEs and ASEs). They are used to reconstruct the length of the day in centennial to millenial timescales and further discussions on the long-term variability of the sea level, global ice amounts, and the core-mantle interactions (Lambeck *et al*., 2014; Stephenson *et al*., 2016; Rekier *et al*., 2022; Hayakawa *et al*., 2025).

Previous studies of TSEs and ASEs focused on their totality/annularity duration and width of the totality/annularity path; however, the lunar limb profiles complicated their analyses (D+80; S+83; RD12; D+16). Chronologically, previous studies mostly examined the eclipse records in the 20th to 21st centuries (S+83; Kubo, 1993; Kilcik *et al*., 2009; Sôma *et al*., 2012, hereafter S+12; Lamy *et al*., 2015; Quaglia *et al*., 2021) but exceptionally extended their coverage to the TSEs in 1715 and 1733 (D+80; Parkinson *et al*., 1980; RD12; D+16; Hayakawa *et al*., 2026a, hereafter H+26a). Early eclipse records are crucial to discussions on whether the long-term $R_{\odot}$ variability showed the secular trends or remained within measurement uncertainties (D+80; Parkinson *et al*., 1980; D+16).

In this context, the scientific community is fortunate to have accesses to Clavius' eclipse reports in 1560 and 1567 (Clavius, 1581). Previous studies have debated whether the 1567 eclipse[3] was total or annular. The TSE in 1567 was a rare hybrid eclipse, meaning that local observations across the eclipse path serve as a valuable key constraint in the variability of the Earth's rotation speed as well as $R_\odot$ at that time (E+80; Stephenson *et al*., 1997, hereafter SJM97; Sigismondi *et al*., 2008, hereafter SBA08; Kilcik *et al*., 2009, hereafter K+09). Each interpretation had some limitations. The TSE hypothesis (assuming the canonical $R_\odot$ value) faced difficulty, as "several beads of photospheric light were visible even where the eclipse was central" (SJM97, p. 347). The ASE hypothesis required the $R_\odot$ value to be 5" larger than the canonical value as well as a substantial shrinkage of the Sun from 1567 to the modern era.

[3] This eclipse was a rare hybrid eclipse: first as annular early in its pass, turning total in the middle, and then becoming annular again late in the path.

However, this discrepancy can be further investigated with the application of recent developments in astronomical datasets. For example, observations taken with satellite missions have greatly improved knowledge of the lunar limb profile (KAGUYA: Araki *et al*., 2009, hereafter A+09); ephemeris data with laser ranging measurements, including NASA JPL DE 441 (Park *et al*., 2021, hereafter P+21); and the variability of the Earth's rotation speed using eclipse and occultation records (Stephenson *et al*., 2016; Morrison *et al*., 2021, hereafter M+21; Hayakawa *et al*., 2026b, hereafter H+26b). Notably, such recent improvements provided by these classic studies have driven renewed interest in eclipse records for $R_\odot$ measurements. As the TSE is so close to the visibility limit in Rome, it can be used to derive the upper $R_\odot$ limit in 1567 that predates the earliest such data in 1715 by almost a century and a half. In this study, we address the discrepancies in the interpretations of eclipse records, documenting our methodology and the modern data used (Section 2). We examine Clavius' observations, revise calculations for the Earth's rotation speed variability in 1560 and 1567 (Section 3), and constrain the $R_\odot$ in 1567 (Section 4) using the ephemeris data of JPL DE 441 (P+21) and the lunar topography data of the KAGUYA mission (A+09).

## 2. Clavius and His Eclipse Observations

Clavius' eclipse reports (Figure 1: Clavius, 1581, pp. 425-426) translate as follows:

ET vt, quod ipſi quoque aliquando obſeruauimus hac in parte, in medium proferamus, recitabo duas inſignes Eclipſes Solis, quæ meo tempore contigerunt non ita pridem, quarum vnam anno 1559. Conimbricæ in Luſitania circa meridiem obſeruaui, in qua interponebatur Luna directe inter viſum, ac Solem, ita vt totum Solem non modico tẽporis interuallo contegeret, eſſentq; tenebræ quodammodo maiores, quàm nocturnæ. Neq; enim, vbi pedem quis poneret, videre poterat, clariſsimeq; ſtellæ in cælo apparebant, & (quod mirabile erat) aues ex aere in terram, præ horrore tam tetræ obſcuritatis, decidebant. Alteram Romæ anno 1567. circa etiam meridiem conſpexi, in qua rurſus Luna etſi inter viſum, ac Solem interijciebatur, non totum tamen Solem obſcurabat, vt in priori, ſed (quod nunquam fortaſsis alias euenit) relinquebatur in Sole circulus quidam exilis circum circa totam Lunam ambiens. Ex quibus duabus duabus Eclipſibus perſpicue admodum colligitur, Solem, & Lunam in vtraque Eclipſi non habuiſſe eandem diſtantiam à terra, vel inter ſe. Si enim eandem diſtantiam & inter ſe, & à terra habuiſſent, quis non videt, eodem modo Solem debuiſſe in vtraque Eclipſi obſcurari? Id quod à Perſpectiuis facile demonſtrabitur, & res perſpicua eſt in manu. Si namque manus eandem ſemper diſtantiam habet à muro aliquo, & ab oculo, ita vt inter murum, & oculum collocetur, perpetuo eandem partem muri è conſpectu auferet, non autem nunc maiorem, & nunc minorem. Igitur nulla ratione dici poteſt, duo hæc luminaria in concentricis orbibus moueri, quia hac ratione ſemper æqualiter inter ſe, & à terra diſtarent; atque adeo apparentia hæc Eclipſium Solarium locum nullo modo poſſet habere.

Figure 1: Clavius' original text from his eclipse reports in 1560 and 1567 (Clavius, 1581, pp. 425–426).

*"And, in order to bring forward what we ourselves have also once observed in this matter, I will recount two notable eclipses of the Sun, which occurred in my own time, not very long ago. One of these I observed around noon in the year 1559 [NB: 1560] at Coimbra in Lusitania, in which the Moon was interposed directly between our sight and the Sun, so that it covered the whole Sun for a considerable interval of time[4], and the darkness was, in a certain sense, deeper than that of night. For one could not see where to place one's foot; very bright stars appeared in the sky; and, what was wonderful, birds fell from the air to the ground out of terror at such dreadful darkness.*

*The other I saw at Rome in the year 1567, also around noon. In this case too, although the Moon was interposed between our sight and the Sun, it did not obscure the whole Sun, as in the former eclipse;*

[4] This is contrasted with "a short interval" in SJM97's translation.

*rather (something that perhaps had never happened before) a certain slender circle was left in the Sun, surrounding the whole Moon on every side".*

The passage is found in the enlarged 1581 edition of his *In sphaeram Ioannis de Sacro Bosco commentarius*[5] and in his subsequent editions, whereas the first edition of the work appeared in 1570 without this passage. Here, Clavius recollected two solar eclipses that he had personally observed: one at Coimbra in Lusitania (modern-day Portugal), which he dated to 1559 and later corrected to 1560[6], and another in Rome in 1567. His description of the first eclipse states that the Moon covered the whole Sun for a considerable interval of time, whereas his account of the second precisely stresses the opposite point: the Sun was not entirely obscured, and "a certain slender circle" remained around the Moon. The corrected chronology accords well with Clavius' known itinerary: he had entered the Jesuit College at Coimbra in 1556, returned to Rome by mid-1561 for theological studies at the Collegio Romano, and began teaching mathematics there by 1563 (Lattis, 1994, pp. 14–15).

We also located Clavius in 1567 at Collegio Romano following SJM97 and corrected the geographical coordinates to N41°53'55", E012°28'49". We located Clavius in 1560 at the Jesuit College, which is locally known as Colégio de Jesus in Coimbra and located at N40°12'36", W008°25'28". This is slightly northward from the neighbourhood of the Colégio São Teotónio (N40°12'00", W008°25'12"), near which SJM97 located the coordinates of his observational site. Qualitatively, Clavius' description of two events may be interpreted as if the lunar disk was much larger than $R_{\odot}$ of 1560 during the eclipse of 1560, while the 1567 eclipse might have been an ASE as suggested by E+80 or a TSE as suggested by SJM97. Below, we critically revisit how the 1567 eclipse was viewed from Clavius' location.

**3. The Earth's Rotation Speed Variability in 1560 and 1567**

Clavius documented and contrasted his experience of the two TSEs on 1560 Aug 21 and 1567 Apr 9. As documented above, Clavius was attached to Colégio de Jesus of the University of Coimbra

[5] This title is translated as "Commentary on the Sphere of Johannes de Sacrobosco".

[6] Clavius later adopted this correction, changing the year to 1560 in the Mainz edition of *In Sphaeram* (Hiraoka, 2005, pp. 104–105). Besides, SJM97 reasonably corrected the date of the first eclipse to 1560 based on their astronomical calculations, identifying this eclipse with a TSE on 1560 Aug 21.

(N40°12'36", W008°25'28") around the 1560 eclipse and Collegio Romano (N41°53'55", E012°28'49") around the 1567 eclipse. M+21's $\Delta T$ spline curve shows the $\Delta T$[7] in 1560 and 1567 as 177 s and 165 s, respectively. These cases show that the maximal magnitudes of Clavius' TSEs were considerably small (Appendix 1). At such a small maximal magnitude, the lunar limb profile may have affected the local eclipse visibility, as the visibility of the photosphere, even of a small portion, can easily form Baily's beads or a diamond ring and change the description of the observer (SJM97; H+26b). Therefore, we used the lunar topography data of the KAGUYA mission (A+09), which was unavailable at the time of SJM97. This dataset is based on satellite measurements and is in substantially better agreement with the lunar occultation observations than Watts' lunar limb profile, which was constructed from ground-based photographs and was widely used prior to the KAGUYA mission (see *e.g.*, Figure 3 of S+12).

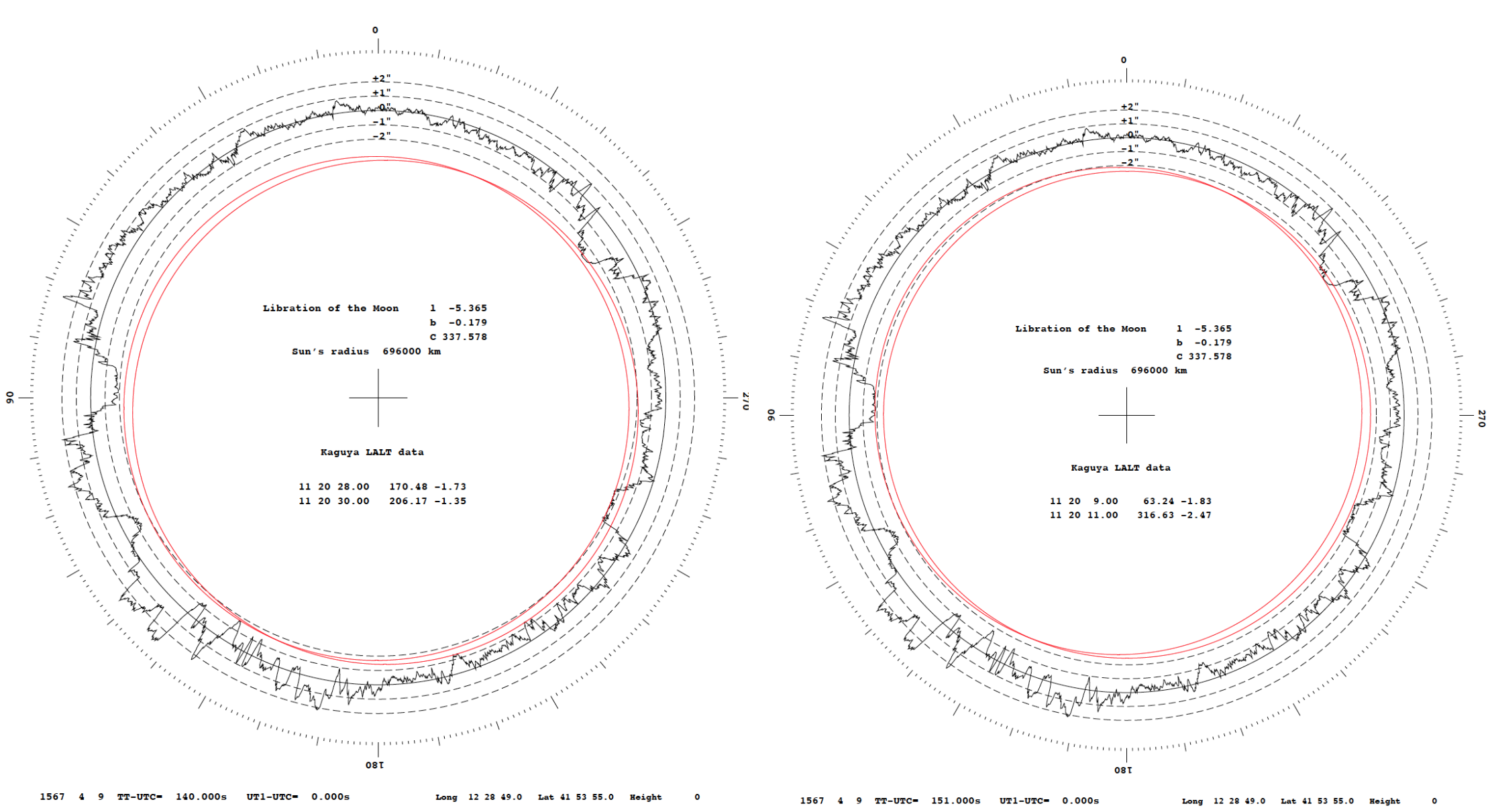

Figure 2: Lunar limb profile in comparison with the solar limb profile under boundary $\Delta T$ values: (left) 140 s and (right) 151 s, when the $R_{\odot}$ is set to the IAU canonical value (696000 km). The position angles are measured from the direction of the celestial north, in contrast with SJM97, who may have obtained the position angle from the lunar rotation axis. The lunar topography data are derived from the KAGUYA mission (A+09), following S+12's method. The lunar limb profile is emphasised by a factor of ≈ 50 for visual presentation (see also H+26a).

[7] $\Delta T$ is a parameter for the variable Earth's rotation. This parameter is defined as $\Delta T = \mathrm{TT} - \mathrm{UT}$, where TT means Terrestrial Time (a theoretical uniform timescale) and UT means Universal Time (time defined by the Earth's rotation).

Figure 2 shows the boundary $\Delta T$ values, where the contemporaneous lunar limb profile managed to completely hide the solar photosphere, using the calculation method of S+12 and assuming the IAU canonical $R_{\odot}$ value (696000 km). In this case, the local totality is satisfied within the $\Delta T$ margins of $-492$ s $\leq \Delta T \leq 200$ s in 1560 and 140 s $\leq \Delta T \leq 151$ s in 1567[8] (Figure 3). In contrast with SJM97, our results allow the lunar body to completely hide the photosphere and make a TSE visible to Clavius in Rome. Our results contrast with M+21's $\Delta T$ margins of $-480$ s $< \Delta T < 210$ s in 1560 and 145 s $< \Delta T <$ 165 s in 1567, with our 1567 margin appreciably lower than M+21's $\Delta T$ spline curve ($\Delta T$ = 165 s). As shown in Figure 3, our results form the basis for modifications on the $\Delta T$ spline curve in the 16th century, in combination with the recent comprehensive modifications of the contemporaneous $\Delta T$ constraints (H+26b). In combination with the 1514 eclipse data (H+26b), our results require a slightly steeper $\Delta T$ decrease between 1514 and 1567 (Figure 3).

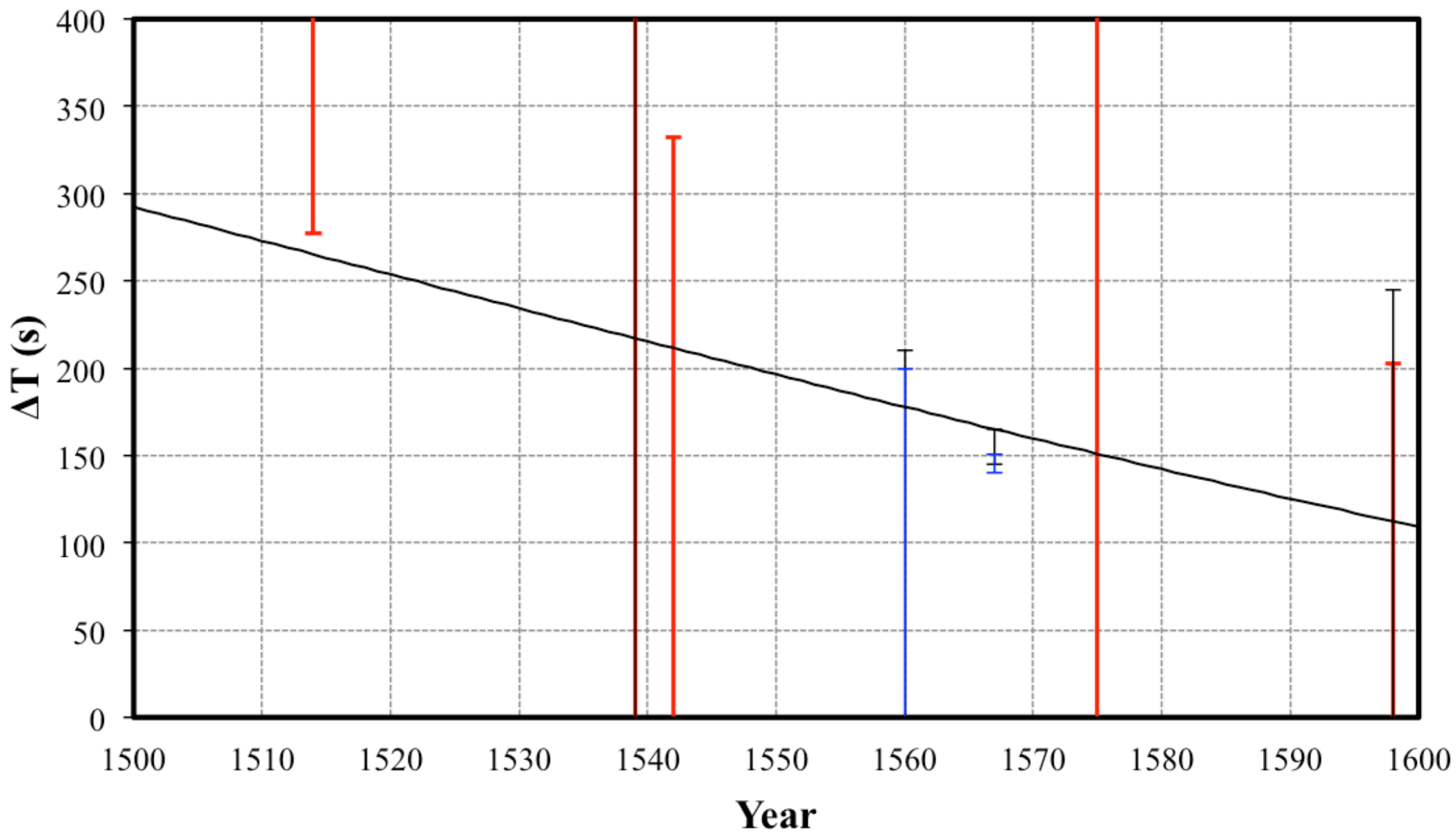

Figure 3: $\Delta T$ constraints derived from Clavius' eclipse reports in this study (blue bars) and those in the previous studies (black bars for M+21 and red bars for H+26b), in comparison with M+21's $\Delta T$ spline curve (black line) .

[8] Our calculations indicate the $\Delta T$ margins as $-493$ s $\leq \Delta T \leq 207$ s for the 1560 eclipse and 113 s $\leq \Delta T \leq 181$ s for the 1567 eclipse to locate these sites in their totality paths if the method of Sôma and Tanikawa (2015) is used, and assuming the IAU canonical $R_{\odot}$ value (696000 km) and a perfectly circular lunar limb.

## 4. $R_\odot$ Measurement Based on the 1567 TSE Report

What Clavius described for the 1567 eclipse is simple. Clavius saw "a certain slender circle" being "left in the Sun" and "surrounding the whole Moon on every side" and considered this as "something that perhaps had never happened before" (Clavius, 1581, pp. 425–426)[9]. Whatever Clavius had in his understanding, this description leaves substantial room for interpretation on this "slender circle", either as annulus upon ASE, an inner corona upon TSE, or a chromosphere upon TSE (Appendix 2).

Clavius' report allows us to constrain the $R_\odot$ value of 1567 based on three scenarios and S+12's methodology. The second and third scenarios allow us to interpret this account as a TSE at Collegio Romano and to estimate the $R_\odot$ value as $R_\odot \leq$ 696200 km in absolute size (959.92" in angular size). We repeated the same calculation described in Figure 2, narrowing down the $\Delta T$ value and increasing the $R_\odot$ value. Figure 4 shows the upper limit of the local TSE visibility at $\Delta T$ = 144 s and $R_\odot$= 696200 km. This threshold puts a hard upper limit for the $R_\odot$ value of 1567. Setting $R_\odot$ > 696200 km, Collegio Romano can see Baily's beads with significant brightness and may have missed such a ring structure either in the inner corona or chromosphere.

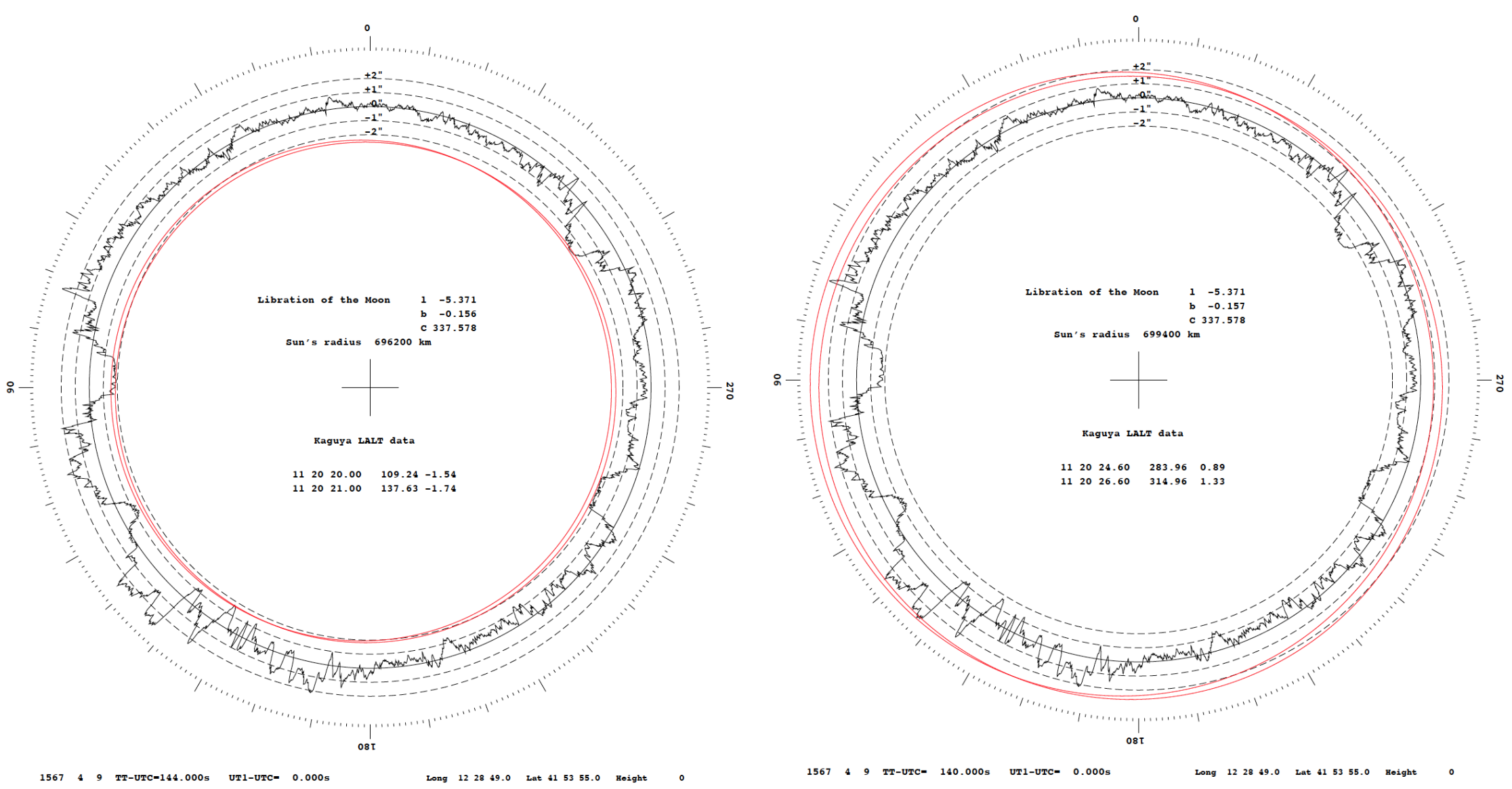

[9] Clavius may have taken this eclipse as a case in which the Moon could not cover the entire Sun in contrast with the TSE of 1560 (Clavius, 1581, pp. 424–426). Owing to this description, some modern scientists associate Clavius' 1567 report with an ASE (*e.g.*, E+80; SBA08).

Figure 4: Lunar limb profiles in comparison with the boundary $R_\odot$ values: (left) 696200 km as an upper $R_\odot$ limit for the local TSE visibility and (right) 699400 km as a lower $R_\odot$ limit for the local ASE visibility, following the procedure of Figure 2. Here, we set their boundary conditions where the solar photosphere touches the bottom of (left) the deepest lunar valleys and (right) the top of the lunar mountains.

Following the first scenario, a local annularity at Collegio Romano and a substantially larger $R_\odot$ value in 1567 of 699400 km ≤ $R_\odot$ in absolute size are required, *i.e.*, 967.13" ≤ $R_\odot$ in angular size. Figure 4 shows the ultimate limit of the ASE visibility at $\Delta T$ = 140 s and $R_\odot$ = 699400 km. Below this $R_\odot$ threshold, the annulus would have been blocked by the lunar limb structure and could not form a complete annulus as a ring structure. This requires an $R_\odot$ excess of 7.5" in angular size against the IAU nominal value, which is substantially larger than SBA08 and K+09's excess estimates (≈ 5.0").

This excursion is too extreme even for E+80's shrinking-Sun hypothesis to satisfy. E+80 estimated the horizontal and vertical $R_\odot$ shrinkage to be 1.125"/century and 0.375"/century, respectively. We have only 3 centuries between Clavius' 1567 eclipse report and the establishment of Auwers' canonical $R_\odot$ value of 959.63" in angular size (Auwers, 1891). Even Eddy's extreme hypothesis can only expand the $R_\odot$ back in 1567 as large as 3.645" and 1.215" in horizontal and vertical $R_\odot$ excesses, respectively. This results is substantially smaller than the requirement of the $R_\odot$ excess of 7.5" for the ASE hypothesis. Therefore, without a more extreme hypothesis, we can confidently reject the ASE hypothesis and set the $R_\odot$ constraint of 1567 as $R_\odot$ ≤ 696200 km in absolute size and $R_\odot$ ≤ 959.92" in angular size.

**6. Summary and Discussions**

This study analysed Clavius' reports of eclipses that occurred in 1560 and 1567 (Figure 1). Clavius explicitly reported the totality of the 1560 eclipse in Coimbra and described a slender circle around the Moon during the 1567 eclipse. Bibliographical studies on Clavius allow us to determine the probable observational sites as Colégio de Jesus in Coimbra (N40°12'36", W008°25'28") for his 1560 eclipse account and Collegio Romano in Rome (N41°53'55", E012°28'49") for his 1567 eclipse account. Some interpret the his account of the 1567 eclipse as a TSE, whilst other interpreted his account as indicating that he saw an ASE.

To investigate whether the 1567 eclipse as seen from Rome was a TSE or an ASE, we computed the variability of the Earth's rotation speed in the 1560s following S+12's method, the ephemeris data of NASA JPL DE 441 (P+21), the lunar topography data of the KAGUYA mission (A+09), and the IAU canonical $R_\odot$ value (696000 km). Our findings revealed that the Moon completely covered the photosphere during the 1567 eclipse, even if the lunar limb profile was considered (Figure 2).

According to our calculations, the local TSE visibility is satisfied within the $\Delta T$ margins of $-492$ s $\leq \Delta T \leq 200$ s in 1560 and 140 s $\leq \Delta T \leq 151$ s in 1567. This result contrasts with that of SJM97, who considered that "the eclipse was neither fully total nor annular" and "several beads of photospheric light would be visible even where the eclipse was central" (SJM97, p. 347). Moreover, our $\Delta T$ constraint in 1567 requires us to revise M+21's $\Delta T$ spline curve downward in 1567 to satisfy this narrow $\Delta T$ margin, indicating a slightly steeper $\Delta T$ decrease between 1514 and 1567 in correlation with H+26b (Figure 3).

This study also attempted to constrain the $R_\odot$ value using 1567 using Clavius' accounts. This study utilised three scenarios to interpret what Clavius reported as a slender circle around the Moon: (1) the annulus of an ASE, (2) the inner corona of a TSE, and (3) the chromosphere of a TSE. Scenario (1) requires the $R_\odot$ to have shrunk 7.5" within 3 centuries between 1567 and Auwers (1891). This is too drastic even in comparison with Eddy's extreme hypothesis of the shrinking Sun.

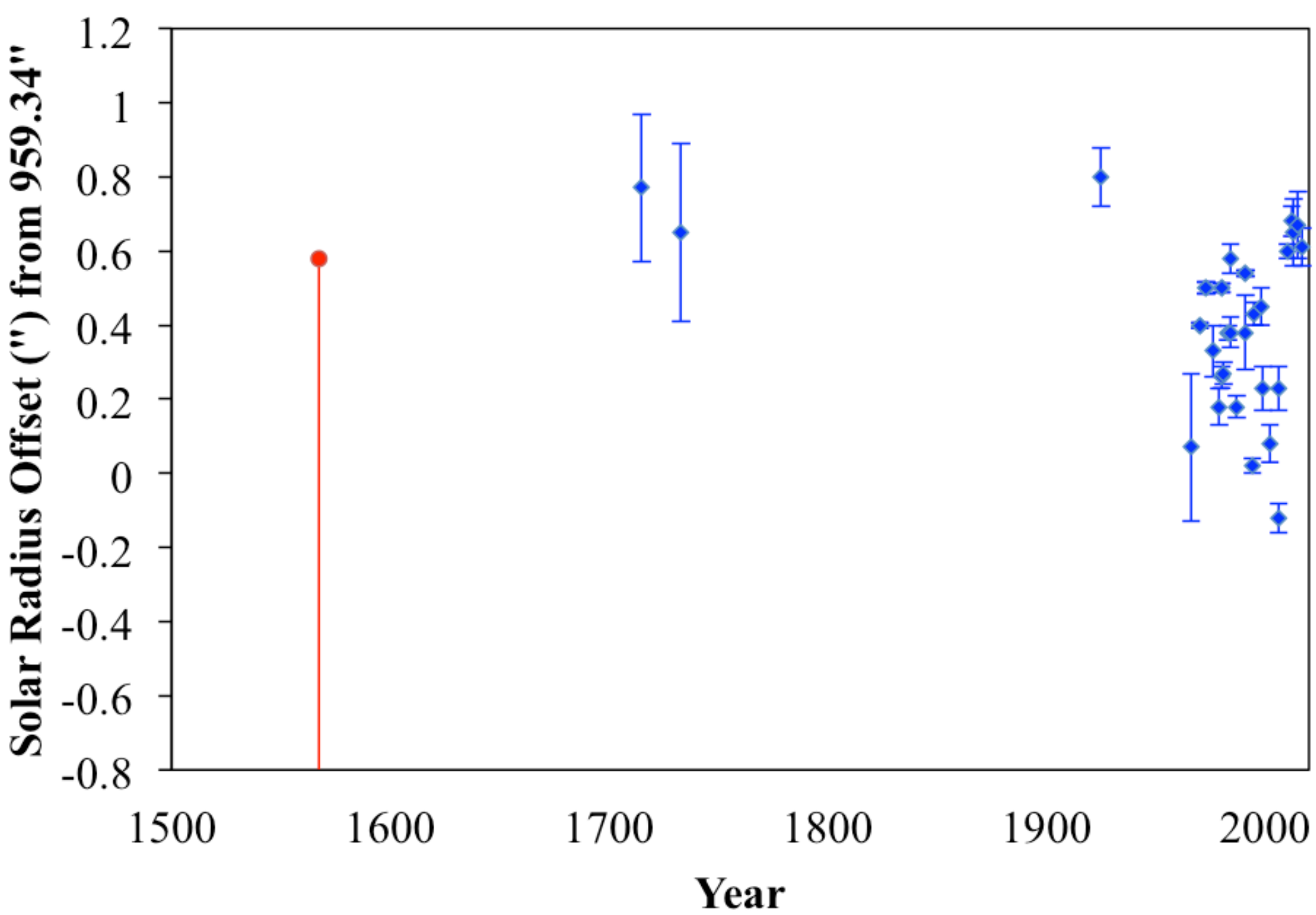

Figure 5: Eclipse-based $R_\odot$ measurements in our study (red bar) and those of previous studies (blue bars), in comparison with the modern helioseismic $R_\odot$ estimate (959.34") of Takata and Gough (2024). The blue bars compositely show the $R_\odot$ estimates of the previous studies (D+80; S+83; Kubo, 1993; K+09; Lamy *et al*., 2015; Quaglia *et al*., 2021; H+26a)[10].

In contrast, Scenarios (2) and (3) indicate a TSE and look far more realistic. In each case, the $R_\odot$ value must remain below $R_\odot \leq 696200$ km in absolute size, *i.e*., $R_\odot \leq 959.92"$ in angular size (Figures 4 and 5). Compared with previous studies, our results, based solely on eclipse-based $R_\odot$ measurements, indicate that the $R_\odot$ value did not show a significant secular shrinkage from 1567 onward. The upper $R_\odot$ limit of 1567 (959.92") compares the lower $R_\odot$ limit of 1715 (959.91") and yields at best a 0.01" reduction during the 148 years in between. The upper $R_\odot$ limit of 1567 (959.92") is also consistent with the 2017 $R_\odot$ constraint (959.95±0.05"). Furthermore, the 1925 $R_\odot$ constraint (960.14±0.08") indicates the lower limit of 960.06" and a slight increase from the 1567 $R_\odot$ upper limit[11]. These cases indicate that the Sun may not have exhibited a secular shrinkage or remained stable but may have shown some $R_\odot$ oscillations in the centennial timescale, in contrast with the findings of previous studies (D+80; Parkinson *et al*., 1980; Hiremath *et al*., 2020). However, to further pursue this possibility, we ultimately need to revisit the best available records with homogeneous methods and datasets, considering the error estimates in an appropriate statistical analysis.

[10] Figure 5 does not show D+16's data, as they have considerable data offsets with their probable source articles (*e.g*., D+80; S+83).
[11] It has been suggested that the $R_\odot$ has an anticorrelation with the solar activity level (Secchi-Rosa Law: Secchi, 1872). In this regard, The International Sunspot Number (Version 2) and $R_\odot$ estimates based on SSN were 9.3±23.5 (vs $R_\odot \leq 959.92"$) in 1567 (Usoskin *et al*., 2021), 8.5±1.0 (vs $R_\odot$ =

**Data Availability**

The source reports studied here are found in Clavius (1581). This document is digitally available thanks to the ETH-Bibliothek Zürich. The ephemeris data are provided from Park *et al*. (2021). The $\Delta T$ spline curve has been provided by Morrison *et al*. (2021). The lunar topography data are provided from the KAGUYA mission (Araki *et al*., 2009).

**Funding**

This research was conducted under the financial support of JSPS Grants-in-Aids JP25K17436 and JP25H00635, the ISEE director's leadership fund for FYs 2021 – 2026, the Young Leader Cultivation (YLC) programme of Nagoya University, Tokai Pathways to Global Excellence (Nagoya University) of the Strategic Professional Development Program for Young Researchers (MEXT), the young researcher units for the advancement of new and undeveloped fields in Nagoya University Program for Research Enhancement, the Interdisciplinary Research Strategy Projects of the Institute for Space−Earth Environmental Research (ISEE), Transdisciplinary Network linking Space-Earth Environmental Science, History, and Archaeology (JPMXP1324134720) of MEXT Promotion of Development of a Joint Usage/Research System Project: Coalition of Universities for Research Excellence Program (CURE), Research Grants in the Humanities of the Mitsubishi Foundation (202520034), and the NIHU Multidisciplinary Collaborative Research Projects NINJAL unit 'Rediscovery of Citizen Science Culture in the Regions and Today'.

---

960.11± 0.20") in 1715 (Carrasco *et al*., 2022), 8.3 (vs $R_\odot$ = 959.34 ± 0.22") in 1733 (Clette *et al*., 2023), 74.0±7.7 (vs $R_\odot$ = 960.14±0.08") in 1925, and 21.7±2.5 (vs $R_\odot$ = 959.95±0.05") in 2017 (Clette *et al*., 2023), respectively (see also Figure 5). Since the activity level in 1925 was significantly higher as compared with TSEs in 1560, 1567, 1715, and 1733, the $R_\odot$ should be expected to be smaller (due to an anticorrelation with the solar activity level), which does not seem the case (see Figure 5). Further investigations are needed for the $R_\odot$ variability in comparison with the level of solar activity.

**Acknowledgments**

Eclipse calculations were partly carried out on the Multi-wavelength Data Analysis System operated by the Astronomy Data Center, National Astronomical Observatory of Japan. HH thanks Marinus van der Sluijs for stimulating discussions. The National Solar Observatory (NSO) is operated by the Association of Universities for Research in Astronomy (AURA), Inc., under cooperative agreement with the National Science Foundation.

**Appendix 1: The Eclipse Magnitude Estimate without Considering the Lunar Limb Profile**

We compute the maximal magnitude of the eclipse in Coimbra in 1560 as 1.002 at 11:48:59 UT (11:15:27 LAT) and that of Rome in 1567 as 1.001 at 11:19:47 UT (12:10:41 LAT), using the method of Sôma and Tanikawa (2015), ephemeris data of the NASA JPL DE 441 (P+21), M+21's $\Delta T$ spline curve ($\Delta T$ = 165 s), and the canonical $R_\odot$ value of the IAU (696000 km) and a perfectly circular lunar limb.

Our results are slightly different from what the interactive map of Espenak and Meeus (2009) shows for the 1560 eclipse and the 1567 eclipse. There are some reasons. First, in this study, we have defined the eclipse magnitude (M) as M = (Sun's apparent angular semidiameter + Moon's apparent angular semidiameter - apparent angular distance of the centers of the Sun and Moon) / (Sun's apparent angular diameter), following H+26a. This definition is the same with what Espenak and Meeus (2009) used for their M upon partial solar eclipses. However, Espenak and Meeus (2009) defined the M upon the TSEs and ASEs differently as M = Moon's apparent angular diameter / Sun's apparent angular diameter. Their two M definitions bring a discontinuity on M between the partiality and totality, make M stable over the totality/annularity phase, obscure the actual magnitude evolution upon the totality/annularity phase, and cause inhomogeneity with their M definition for lunar eclipses.

Second, we follow the IAU Resolution of 1982 to use the same lunar diameter, occasionally taking the lunar topography data into consideration. Third, Espenak and Meeus (2009) uses earlier ephemeris data (ELP2000/82 for the Moon and VSOP87 for the Sun) and older $\Delta T$ spline curve (Morrison and Stephenson, 2004), while we are using the ephemeris data of JPL DE441 (P+21) and the M+21's $\Delta T$ spline curve here.

**Appendix 2: Circular Structure around the Moon upon the TSE on 1567 Apr 9**

First, some authors interpreted this ring as the annulus of an ASE and required a substantially larger $R_\odot$ value (E+80; Lattis, 1994; SBA08; K+09), probably owing to the contrast with Clavius' 1560 eclipse account. Conversely, from as early as Kepler to modern scholarship, various possibilities other than an interpretation of Clavius' account of an ASE have been suggested (Hammer, 1939, pp. 257–262). This possibility is discussed in the main text.

Second, SJM97 associated this ring with an inner corona. Initially, their interpretation was inconsistent with that of the Moon with Baily's beads, as described by SJM79, because Baily's beads (i.e., a photosphere with a fragmental visibility) were too bright to keep the "circle" structure around the Moon, overshining the inner corona (Littmann *et al*., 2008, p. 189) and altering the impression of the observer. Our calculations (Figure 2) made this scenario more plausible by allowing the Moon to hide the entire photosphere from Collegio Romano. This hypothesis sounds attractive, as the inner corona looks much brighter near the edge of the totality path (Littmann *et al*., 2008, pp. 136–137).

This interpretation is challenged by the alleged absence of this structure in the 1560 eclipse. The inner corona spreads approximately 0.2–0.3 $R_\odot$ out of the photosphere (DeForest, 2007) and cannot be hidden with the maximal magnitude of the 1560 eclipse (≈ 1.002). Morphologically, this circular structure is inconsistent with that of coronal streamers. While the lesson of the Maunder Minimum (Hayakawa *et al*., 2021) generally indicates their possible loss in the grand minima, the 1560 eclipse occurred following the end of the Spörer Minimum in the early to mid 16th century (Usoskin *et al*., 2021).

Third, this circular structure may be associated with the chromosphere. The chromosphere extends up to 1000–5000 km above the photosphere (Figures 2–4 of Carlsson *et al*. (2019)). This is too low to have been seen in the 1560 eclipse but may have been sufficiently high to enable visibility in a marginal case like the 1567 eclipse.